\documentclass[aps,prl,reprint,superscriptaddress,showpacs]{revtex4-1}

\usepackage{amsmath}
	\usepackage{makeidx}
	\usepackage{amsfonts}
	\usepackage[ansinew]{inputenc}
	\usepackage[usenames,dvipsnames]{pstricks}
	\usepackage{subfigure}
	\usepackage{epsfig}
	\usepackage{pst-grad} 
	\usepackage{pst-plot} 
	\usepackage{mathrsfs}
	\usepackage{graphicx}
  \usepackage{bm}
  \usepackage{amssymb}
\usepackage[colorlinks=true,linkcolor=blue]{hyperref}
\expandafter\ifx\csname package@font\endcsname\relax\else
 \expandafter\expandafter
 \expandafter\usepackage
 \expandafter\expandafter
 \expandafter{\csname package@font\endcsname}
\fi
\makeindex
\begin{document}

\title{Nonlinear Helicons ---an analytical solution elucidating multi-scale structure}

\author{Hamdi M. Abdelhamid}
\email{hamdi@ppl.k.u-tokyo.ac.jp}
\affiliation{Graduate School of Frontier Sciences, The University of Tokyo, Kashiwanoha, Kashiwa, Chiba 277-8561, Japan}
\affiliation{Physics Department, Faculty of Science, Mansoura University, Mansoura 35516, Egypt}
\author{Zensho Yoshida}
\email{yoshida@k.u-tokyo.ac.jp}
\affiliation{Graduate School of Frontier Sciences, The University of Tokyo, Kashiwanoha, Kashiwa, Chiba 277-8561, Japan}
\begin{abstract}
The helicon waves exhibit varying characters depending on plasma parameters, geometry, and wave numbers.
Here we elucidate an intrinsic multi-scale property embodied by the combination of dispersive effect and nonlinearity.
The extended magnetohydrodynamics model (exMHD) is capable of describing wide range of parameter space.
By using the underlying Hamiltonian structure of exMHD, we construct an exact nonlinear solution
which turns out to be a combination of two distinct modes, the helicon and Trivelpiece-Gould (TG) waves.
In the regime of relatively low frequency or high density, 
however, the combination is made of the TG mode and an ion cyclotron wave (slow wave).
The energy partition between these modes is determined by the helicities carried by the wave fields. 
\end{abstract}
\maketitle

The helicons (synonymously-called whistlers) have variety of applications
such as plasma sources\cite{Boswell1984,Shinohara2009,Chent, Weitang}, spacecraft propulsion\cite{Toki2006,Franklin2006, Ahedo2011, Squire} as well as in laboratory plasma experiments \cite{Takahashi2016,Stenzel2015a,Takahashi2015,Takahashi2011}.
Helicons are low frequency (compared with the electron cyclotron frequency) circularly polarized electromagnetic waves propagating along an ambient magnetic field. 
The linear theory of helicon waves has been studied in great detail for a particular frequency range much lower than electron cyclotron frequency and much higher than the ion cyclotron frequency, in which the ions are considered immobile (see \cite{Boswell1997, Chen1997a}, and references therein). 
While early theory ignored electron mass, 
Boswell\cite{Boswell72} found that a finite electron inertia gives rise to a second quasi-electrostatic wave called Trivelpiece-Gould (TG) wave\,\cite{Trivelpiece1959}. 

In this Letter, we present an analytical nonlinear helicon wave satisfying an extended MHD (exMHD) system
which takes into account the two-fluid effects due to the electron inertia and Hall drift
\cite{Abdelhamid2015a, KeramidasCharidakos2014, Goedbloed2004}:
\begin{eqnarray}
\label{conteq}
\frac{\partial \rho}{\partial t}=-\nabla\cdot\left(\rho\bm{V}\right),
\end{eqnarray}
\begin{eqnarray}
\label{momeq}
\frac{\partial \bm{V}}{\partial t}&=&-\left(\nabla\times\bm{V}\right)\times\bm{V}+\rho^{-1} \left(\nabla\times\bm{B}\right)\times\bm{B}^{\ast}\nonumber \\ &&-\nabla\left(h+\frac{V^{2}}{2}+d^{2}_{e}\frac{\left(\nabla\times\bm{B}\right)^{2}}{2\rho^{2}}\right),
\end{eqnarray}
\begin{eqnarray}
\label{OhmXMHD}
\frac{\partial \bm{B}^{\ast}}{\partial t}&=&\nabla\times\left(\bm{V}\times\bm{B}^{\ast}\right)- \nabla\times\left(\rho^{-1} \left(\nabla\times\bm{B}\right)\times\bm{B}^{\ast}\right)\nonumber\\&&+d^{2}_{e} \nabla\times\left(\rho^{-1} \left(\nabla\times\bm{B}\right)\times\left(\nabla\times\bm{V}\right)\right),
\end{eqnarray}
where
\begin{eqnarray}
\label{extB}
\bm{B}^{\ast}=\bm{B}+d^{2}_{e}\nabla\times\rho^{-1}\left(\nabla\times\bm{B}\right),
\end{eqnarray}
is a generalized magnetic field;
as to be shown in the expression (\ref{Energy}) of the energy, 
we may regard $\bm{B}^{\ast}$ as the \emph{magnetic field intensity}
(not only the energy but also other important quantities, like helicities, pertain to this $\bm{B}^{\ast}$).
We have used standard notation:
$\rho$ is the mass density, ${\bm V}$ is the center-of-mass velocity, $\bm{B}$ is the magnetic field, $h$ is the total enthalpy, $d_{i\left(e\right)} = c/\left(\omega_{pi\left(pe\right)}L\right)$ is the normalized ion (electron) skin depth, $\omega_{pi\left(pe\right)}=\sqrt{n_{0} e^{2}/\epsilon_{0} m_{i\left(e\right)}} $ is the ion (electron) plasma frequency, $n_{0}$ is a constant density, $c$ is the speed of light and $L$ is the system length scale. 
Note that the above expressions have been normalized in the standard Alfv\'en units, with the magnetic field normalized to an ambient magnetic field $B_{0}$ and the velocity normalized to the Alfv\'en speed $V_{A}=B_{0}/\sqrt{\mu_{0}\rho_{0}}$.

The exMHD system is endowed with a Hamiltonian structure\,\cite{Abdelhamid2015a}. 
On the phase space of state variables $u=(\rho,\bm{V},\bm{B}^*)$, we can define a Poisson bracket,
which has three independent Casimir invariants
(existence of Casimir invariants make the Poisson bracket \emph{noncanonical}):
\begin{eqnarray}
\label{c1}
C_{1}&=&\frac{1}{2}\int_{\Omega}\left(\bm{A}^{\ast}-\frac{2 d^{2}_{e}}{d_{i}}\bm{V}\right)\cdot\bm{B}^{\ast} d^{3}x,
\end{eqnarray}
\begin{eqnarray}
\label{c2}
C_{2}=\frac{1}{2}\int_{\Omega}\Big[&&\left(\bm{A}^{\ast}+d_{i}\bm{V}\right)\cdot\left(\bm{B}^{\ast}+d_{i}\nabla\times\bm{V}\right)\nonumber\\&&+d^{2}_{e}\bm{V}\cdot\left(\nabla\times\bm{V}\right)\Big] d^{3}x,
\end{eqnarray}
\begin{eqnarray}
\label{c3}
C_{3}&=&\int_{\Omega}\rho~d^{3}x.
\end{eqnarray}
The energy is given by\,\citep{Kimura,KeramidasCharidakos2014,Abdelhamid2015a}
\begin{eqnarray}
\label{Energy}
E=\int_{\Omega}\left\{\rho\left(\frac{\left|\bm{V}\right|^{2}}{2}+U\left(\rho\right)\right)+\frac{\bm{B}\cdot\bm{B}^{\ast}}{2}
\right\} d^{3}x.
\end{eqnarray}
Writing $E$ in terms of $u$ gives the Hamiltonian.

We can invoke the method of \cite{Yoshida12,Yoshidax} to construct nonlinear wave solutions by the Casimir invariants and the Hamiltonian.
We first construct energy-Casimir equilibrium by extremizing
\begin{eqnarray}
\label{ec}
\mathscr{E}_{\mu}\left(u\right)=E\left(u\right)-\sum^{3}_{n=1}\mu_{n} C_{n}\left(u\right).
\end{eqnarray}
The Euler-Lagrange equation $\partial_{u}\mathscr{E}_{\mu}=0$ reads
\begin{eqnarray}
\label{ec1}
\nabla\times\bm{B}=\left(\mu_{1}+\mu_{2}\right)\bm{B}^{\ast}+\left(d_{i}\mu_{2}-\frac{d^{2}_{e}}{d_{i}}\mu_{1}\right)\nabla\times\bm{V},
\end{eqnarray}
\begin{eqnarray}
\label{ec2}
\rho\bm{V}=\left(d_{i}\mu_{2}-\frac{d^{2}_{e}}{d_{i}}\mu_{1}\right)\bm{B}^{\ast}+\left(d^{2}_{i}+d^{2}_{e}\right)\mu_{2}\nabla\times\bm{V},
\end{eqnarray}
\begin{eqnarray}
\label{ec3}
\frac{V^{2}}{2}+h\left(\rho\right)+d^{2}_{e}\frac{\left(\nabla\times\bm{B}\right)^{2}}{2\rho^{2}}-\mu_{3}=0,
\end{eqnarray}
where $\mu_{1}$, $\mu_{2}$ and $\mu_{3}$ are Lagrange multipliers. 
Assuming an incompressible flow $\left(\nabla\cdot\bm{V}=0\right)$ with a constant mass density $\rho=1$, 
equations \eqref{ec1} and \eqref{ec2} combine to yield a triple-curl Beltrami equation
\begin{eqnarray}
\label{3curl}
\nabla\times\nabla\times\nabla\times\bm{B}-\alpha_{1}\nabla\times\nabla\times\bm{B}+\alpha_{2}\nabla\times\bm{B}-\alpha_{3}\bm{B}=0,\nonumber\\
\end{eqnarray}
 where \begin{eqnarray*}
\alpha_{1}&=&\big[\left(d^{2}_{i}+d^{2}_{e}\right)\mu_{2}+d^{2}_{e}\left(\mu_{1}+\mu_{2}\right)\big]/\Delta,
\\
\alpha_{2}&=&\left[1+\left(d^{2}_{i}+d^{2}_{e}\right)\left(\mu_{1}+\mu_{2}\right)\mu_{2}-\left(d_{i}\mu_{2}-\frac{d^{2}_{e}}{d_{i}}\mu_{1}\right)^{2}\right]/\Delta,
\\
\alpha_{3}&=&\big[\mu_{1}+\mu_{2}\big]/\Delta,
\\
\Delta&=&d^{2}_{e}\left[\left(d^{2}_{i}+d^{2}_{e}\right)\mu_{2}\left(\mu_{1}+\mu_{2}\right)-\left(d_{i}\mu_{2}-\frac{d^{2}_{e}}{d_{i}}\mu_{1}\right)^{2}\right].
\end{eqnarray*}
Equation \eqref{3curl} can be factored as
\begin{eqnarray}
\label{3curl2}
\left(curl-\lambda_{0}\right)\left(curl-\lambda_{1}\right)\left(curl-\lambda_{2}\right)\bm{B}=0,
\end{eqnarray}
where the eigenvalues $\lambda_{0}$, $\lambda_{1}$ and $\lambda_{2}$ are given by
\begin{align}
\begin{split}
\label{eigen}
\lambda_{0}+\lambda_{1}+\lambda_{2}&=\alpha_{1},\\
\lambda_{0}\lambda_{1}+\lambda_{1}\lambda_{2}+\lambda_{2}\lambda_{0}&=\alpha_{2},\\
\lambda_{0}\lambda_{1}\lambda_{2}&=\alpha_{3}.
\end{split}
\end{align}
The general solution of \eqref{3curl2} can be written as a linear combination of three Beltrami eigenfunctions\,\cite{Yoshida99m}:
\begin{eqnarray}
\label{gsolb}
\bm{B}=\sum^{2}_{l=0}a_{l}\bm{G}_{l},
\end{eqnarray}
where $\bm{G}_{l}$'s are the Beltrami eigenfunctions satisfying 
$\left(\nabla\times\bm{G}_{l}=\lambda_{l}\bm{G}_{l}\right)$, and $a_{l}$'s are arbitrary constants. 
By substituting \eqref{gsolb} into \eqref{ec1} and \eqref{ec2}, the corresponding flow is given by
\begin{eqnarray}
\label{gsolv}
\bm{V}&=&\sum^{2}_{l=0}\left[\sigma\left(1+d^{2}_{e}\lambda^{2}_{l}\right)+\frac{\left(d^{2}_{i}+d^{2}_{e}\right)\mu_{2}}{\left(d_{i}\mu_{2}-\frac{d^{2}_{e}}{d_{i}}\mu_{1}\right)}\lambda_{l}\right]a_{l}\bm{G}_{l},
\end{eqnarray}
where $\sigma=\left(d_{i}\mu_{2}-\frac{d^{2}_{e}}{d_{i}}\mu_{1}\right)-\frac{\left(d^{2}_{i}+d^{2}_{e}\right)\left(\mu_{1}+\mu_{2}\right)\mu_{2}}{\left(d_{i}\mu_{2}-\frac{d^{2}_{e}}{d_{i}}\mu_{1}\right)}$.
\begin{figure*}[!t]
  \centering
	  \includegraphics[width=0.99\textwidth]{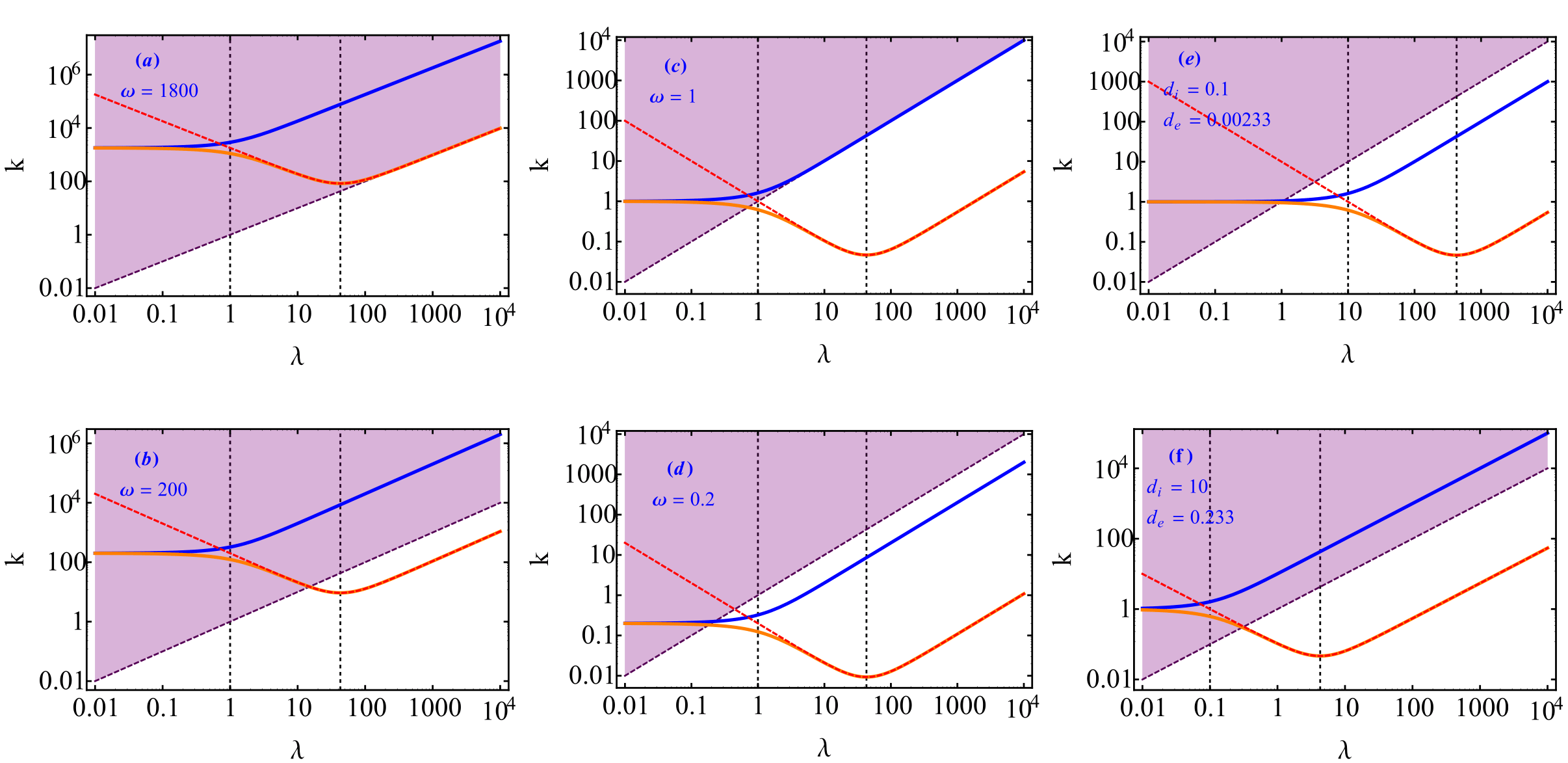}
\scriptsize
\caption{
(Color online)
The relation between $k$ (axial wave number) and $\lambda$ (Beltrami eigenvalue measuring the reciprocal length scale of wave field variation).
In (a), (b), (c) and (d), $d_i=1$ and $d_e=0.0233$ are fixed, while $\omega$ is changed as a parameter.
In (e) and (f), $\omega=1$ is fixed, while the skin depths are changed. 
The shaded region above the dashed line of $k=\lambda$ is the evanescent domain.
The dashed red curve shows the limit of immobile ions given by \cite{Chen1997a}. 
The regime of $\lambda<1/d_i$ may be approximated ky ideal MHD, and the regime of $1/d_i<\lambda< 1/d_e$ by Hall MHD,
while the electron inertia plays important role in the regime of $1/d_e < \lambda$.
}
\normalsize
\label{Fig1}
\end{figure*}
We make a special choice for one of the Beltrami eigenvalues to set $\lambda_{0}=0$.
Then, the corresponding eigenfunction becomes a harmonic field.
Two consequences immediately follow from \eqref{eigen}:
\begin{align}
\mu_{1}=-\mu_{2}&=\mu,\\
\begin{split}
\label{eigen2}
\lambda_{1}+\lambda_{2}&=\alpha_{1},\\
\lambda_{1}\lambda_{2}&=\alpha_{2}.
\end{split}
\end{align}
Now, solving \eqref{eigen2} yields
\begin{eqnarray}
\label{lambda}
\lambda_{1,2}=\frac{d_{i}}{2 d^{2}_{e}\mu_{\star}}\left[1\mp\sqrt{1-\frac{4d^{2}_{e}}{d^{2}_{i}}\left(\mu_{\star}^{2}-1\right)}\right],
\end{eqnarray}
where $\mu_{\star}= \left(d^{2}_{i}+d^{2}_{e}\right)\mu/d_{i}$.
Under this condition the general flow solution becomes
\begin{eqnarray}
\label{gsolv2}
\bm{V}=-\mu_{\star} a_{0}\bm{G}_{0}-\frac{1}{\mu_{\star}}\left(a_{1}\bm{G}_{1}+a_{2}\bm{G}_{2}\right).
\end{eqnarray}

We can write down the Beltrami solutions explicitly, for example, in the cylindrical coordinates:
\begin{eqnarray}
\label{BS}
\bm{B}&=&a_{0}\bm{G}_{0}+\bm{b},~~~~~~~~~\bm{V}=-\mu_{\star}a_{0}\bm{G}_{0}+\bm{v},
\\
\label{bvRel}
\bm{v}&=&\frac{-1}{\mu_{\star}}\bm{b}.
\end{eqnarray}
where 
\begin{eqnarray}
\label{bform}
\bm{b}&=&a_{1}\bm{G}_{1}+a_{2}\bm{G}_{2}\nonumber\\&=&\Bigg\{\sum^{2}_{l=1}a_{l}\left(\lambda_{l}\frac{i m}{r}J_{m}\left(\gamma_{l} r\right)+ik \frac{\partial }{\partial r}\left(J_{m}\left(\gamma_{l} r\right)\right)\right)\hat{r}\nonumber\\&&-\sum^{2}_{l=1}a_{l}\left(\frac{m k}{r}J_{m}\left(\gamma_{l} r\right)+\lambda_{l}\frac{\partial }{\partial r}\left(J_{m}\left(\gamma_{l} r\right)\right)\right)\hat{\theta}\nonumber\\&&+\sum^{2}_{l=1}a_{l}\gamma^{2}_{l}J_{m}\left(\gamma_{l} r\right)\hat{z}\Bigg\}e^{i\left(m\theta+k z\right)},
\end{eqnarray}
where
$\gamma^{2}_{l}=\lambda^{2}_{l}-k^{2}$ measures the transverse wave numbers, $k$ is the axial wave number
and $J_{m}$ is the Bessel function of first kind of order $m$. 


We can derive wave solutions from the forgoing equilibrium solutions.
Evidently \eqref{BS}-\eqref{bform} are equilibrium solutions satisfying
\begin{eqnarray}
\label{inc1}
0&=&\nabla\times\big[\left(\bm{V}-d_{i}\nabla\times\bm{B}\right)\times\bm{B}^{\ast}\big],
\\
\label{inc2}
0&=&\nabla\times\big[\bm{V}\times\left(\bm{B}^{\ast}+\nabla\times\bm{V}\right)\nonumber\\&&+\left(1-d_{i}\right)\left(\nabla\times\bm{B}\right)\times\bm{B}^{\ast}\big],
\end{eqnarray}
\begin{eqnarray}
\label{inc3}
\nabla\cdot\bm{V}&=&0,
\\
\label{inc4}
\nabla\cdot\bm{B}&=&0,~~~~~~~~~\nabla\cdot\bm{B}^{\ast}=0.
\end{eqnarray}
Here we assume that the harmonic field $\bm{G}_{0}$ represents the ambient magnetic field, and the other components $\bm{G}_{1}$ and $\bm{G}_{2}$ are ``wave fields'' propagating on 
 $\bm{G}_{0}$. 
Setting $\left(\bm{G}_{0}=\hat{z}\right)$ and $\left(a_{0}=1\right)$, we write  
\begin{eqnarray}
\label{BS1}
\bm{B}&=&\hat{z}+\bm{b},~~~~~~~~~\bm{V}=-\mu_{\star}\hat{z}+\bm{v},
\end{eqnarray}
Next, we transform the coordinates by Galilean-boost:
\begin{eqnarray*}
\label{trans}
\left(r,\theta,z\right)\longmapsto\left(r,\theta ,\xi\right):=\left(r,\theta,z+\mu_{\star} t\right),
\end{eqnarray*}
where $t\mapsto\tau:=t$ and $z\mapsto\xi:=z+\mu_{\star} t$. The derivatives transform as $\nabla_{r,\theta,z}\mapsto\widetilde{\nabla}_{r,\theta,\xi}$, and $\frac{\partial}{\partial t}\mapsto\frac{\partial}{\partial\tau}+\mu_{\star}\frac{\partial}{\partial\xi}$.
For a 3-vector $\bm{R}$ such that $\nabla\cdot\bm{R}=0$, we may calculate $-\mu_{\star}\frac{\partial\bm{R}}{\partial\xi}=\nabla\times\left(\mu_{\star}\widehat{\bm{e}}_{z}\times\bm{R}\right)$.
Applying the above coordinates transformations along with \eqref{BS1}, the equilibrium equations \eqref{inc1} and \eqref{inc2} transform into
\begin{eqnarray}
\label{inc5}
\frac{\partial \bm{B}^{\ast}}{\partial \tau}=\widetilde{\nabla }\times&\bigg[&\left(\bm{v}-d_{i}\widetilde{\nabla}\times\bm{B}\right)\times\bm{B}^{\ast}\bigg] ,
\\
\label{inc6}
\frac{\partial \left(\bm{B}^{\ast}+\widetilde{\nabla}\times\bm{v}\right)}{\partial \tau}&=&\widetilde{\nabla}\times\bigg[\bm{v}\times\left(\bm{B}^{\ast}+\widetilde{\nabla}\times\bm{v}\right)\nonumber\\&&+\left(1-d_{i}\right)\left(\widetilde{\nabla}\times\bm{B}\right)\times\bm{B}^{\ast}\bigg],
\end{eqnarray}
which read as wave equations. 
Hence, our triple Beltrami solution, which is now denoted as \eqref{BS1},
can be regarded as a wave solution propagating on the ambient field $\bm{B}_{0}=\hat{z}$.

Let us extract wave characteristic quantities from the Beltrami solutions;
we put
\begin{eqnarray}
\label{Bwave}
\bm{B}&=&\hat{z}+\bm{b}~e^{-i\omega t},~~~~~~~~~\bm{V}=\bm{v}~e^{-i\omega t},
\\
\label{bvwaveRel}
\bm{v}&=&\left(\frac{-k}{\omega}\right)\bm{b},
\end{eqnarray}
where $\bm{b}$ is given by \eqref{bform}. 
Equation \eqref{bvwaveRel} determines the relation between the the magnetic and velocity fields in the wave. 
Note that here $\mu_{\star}$ serves as the phase velocity. 
We can rewrite the eigenvalues equation \eqref{lambda} in terms of the frequency $\left(\omega=k\mu_{\star}\right)$ as
\begin{eqnarray}
\label{lambdaw}
\lambda_{1,2}=\frac{d_{i}k}{2 d^{2}_{e}\omega}\left[1\mp\sqrt{1-4\frac{d^{2}_{e}}{d^{2}_{i}}\left(\frac{\omega^{2}}{k^{2}}-1\right)}\right].
\end{eqnarray}
The eigenvalues \eqref{lambdaw} are identical to the eigenvalues obtained from the linear analysis (when the ions are considered immobile), except for the last term under the square-root. 
This term recovers the ion inertia effect, yielding an ion cyclotron wave mode (cf. Fig.\ref{Fig1}) which has been ignored in previous studies.
In Fig.\ref{Fig1} $k-\lambda$ diagrams are presented in which the helicon and TG modes are represented by an ''orange line'' and the ion cyclotron mode by a ''blue line''. We have inserted two vertical dashed lines in each plot to separate the ideal $\left(\lambda<1/d_i\right)$, Hall $\left(1/d_i<\lambda<1/d_e\right)$ and the electron inertia $\left(\lambda>1/d_e\right)$ domains. Notice that the ion cyclotron mode arises when $\lambda$ or $k$ is negative. Figs. \ref{Fig1}(a)-\ref{Fig1}(d) are the $k-\lambda$ curves for different values of the applied magnetic field. We can observe that for $\omega>1800$ in normalized units all modes become evanescent (no propagation). As the value of $\omega$ decreases the helicon and TG modes appeared, when it reaches a definite value ($\omega\leq 1$ in normalized units) a third mode raise. We also observed a particular coupling between the different modes, which for the values of $k<\omega$ the coupling occurs between Helicon and TG waves, whilst the ion cyclotron wave is coupled with the TG wave for the values of $k>\omega$.
To examine the effect of the plasma density,
we plot the $k-\lambda$ relation for two different values of the skin depth;
see. Figs. \ref{Fig1}(e) and \ref{Fig1}(f)). 
The change of the plasma density (thus, the skin depth) strongly influences the waves coupling.
\begin{figure}[!t]
  \centering
	  \includegraphics[width=0.45\textwidth]{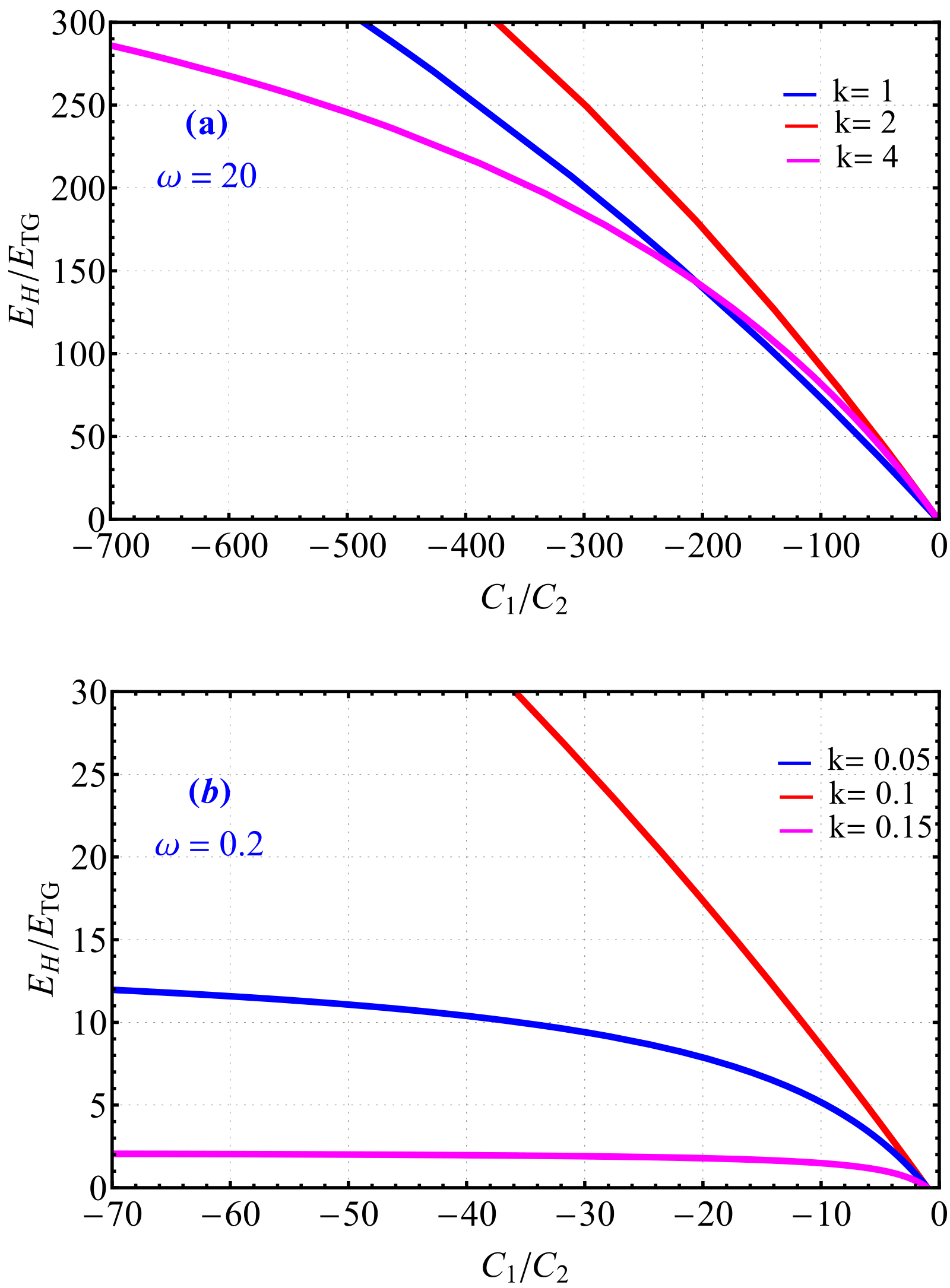}
\scriptsize
\caption{(Color online) The relation between Helicon-TG energy ratio $E_{H}/E_{TG}$ and the helicities ratio $C_{1}/C_{2}$.
Here, we assume $d_i=1$, $d_e=0.0233$ and $m=0$.
In (a) $\omega=20$, and (b) $\omega=0.2$.
}
\normalsize
\label{Fig2}
\end{figure}

As an interesting property of the nonlinear solution \eqref{gsolb}, the
coefficients $a_1$ and $a_2$ can be arbitrarily chosen
to combine two Beltrami eigenfunctions.
Here we study how they are determined by physical conditions.
One possible, conventional argument is to relate such coefficients to boundary conditions on the electromagnetic field.
However, it is known that boundary conditions falls short of determining the coefficients
(in the linear theory)\,\cite{Chen1996, Davies1969, Boswell1984a}.
We can approach the problem from a different angle; based on our Hamiltonian formalism, we can invoke the Casimir invariants (helicities) to quantify the coefficients.
Inserting the solution \eqref{Bwave}-\eqref{bvwaveRel} into \eqref{c1} and \eqref{c2}, 
we obtain the helicities represented in terms of $a_2$ and $a_2$ (and other plasma parameters).
Numerically inverting this somewhat involved relation,
we obtain the ratio $a_1/a_2$, as well as the energy partition, as functions of the helicities. 
Figure \ref{Fig2} shows how the helicon-TG energy ratio is changed by the helicities.
The Casimir invariants \eqref{c1} and \eqref{c2} can be related to the two-fluid (electrons and ions) helicities, respectively, in the one-fluid model limits. 
In two-fluid plasma consisting of electrons and ions, the invariants can be written as $\int_{\Omega} \bm{P}_{e,i}.\nabla\times\bm{P}_{e,i} d^{3}x$, where $\bm{P}_{e,i}$ is the canonical momentum for each specie ($\bm{P}_{e,i}=m_{e,i} \bm{V}_{e,i}+ q_{e,i} \bm{A}$). 
At the one-fluid limit, the exMHD invariants and the two-fluid helicities reduce into the same quantities.
Since the helicity is the measure of the wave polarization, or the twist of the perturbed (generalized) magnetic field lines,
this figure can be used for the practical purpose of designing the wave-launching system to optimize the wave energy partition.

Using the generalized MHD model, we have elucidated the multi-scale structure of electromagnetic waves.
The derived analytical solution, satisfying the set of nonlinear equations,
manifests the intrinsic coupling of the large scale and the electron skin-depth small scale;
the former is realized as a helicon, and the latter as a TG mode.
When the density is sufficiently high or the frequency is low,
the TG mode chooses a different partner, which is the ion-cyclotron slow wave.
It is remarkable that the coupling of such two-modes imitates a linear combination.
Moreover, the dispersion relations obtained by the linearized model do apply for fully nonlinear solutions with arbitrary amplitudes.
This 'superficial linearity' is a manifestation of the beautiful algebraic structure underlying the generalized MHD system.
Such simplicity is due to the fact that the Hamiltonian and the Casimir invariants are quadratic functionals of the wave fields.
For compressible modes, however, the thermal and electrostatic energies add a different type of nonlinearity
which causes solitary-wave like structures \cite{Yoshida12, Emoto14}.


\begin{acknowledgments}
HMA would like to thank the Egyptian Ministry of Higher Education for supporting his research activities. 
The work of ZY was supported by JSPS KAKENHI Grant Number 15K13532, 
\end{acknowledgments}

\bibliographystyle{apsrev4-1}
\bibliography{Helicon}

\end{document}